\documentclass[11pt]{article}
\input epsf
\def\lvm{\leavevmode\hbox to\parindent{\hfill}}

\def\fun#1#2{\lower3.6pt\vbox{\baselineskip0pt\lineskip.9pt
\ialign{$\mathsurround=0pt#1\hfil##\hfil$\crcr#2\crcr\sim\crcr}}}
\def\plotone#1{\epsfysize=15cm\epsffile{#1}}

\def\msun{M_\odot}
\def\Msun{$\msun$}

\def\fefsx{$^{56}$Fe}
\def\cofsx{$^{56}$Co}
\def\nifsx{$^{56}$Ni}

\def\etal{et al. }
\def\sn{SNe~Ia}
 
\title{Thermonuclear Burning Regimes and the Use of SNe~Ia
       in Cosmology}
 
\author{E.I.Sorokina${}^1$, S.I.Blinnikov${}^{2,3}$, O.S.Bartunov${}^1$\\}
\date{}
 
\begin{document}
\begin{titlepage}
 
\maketitle
\thispagestyle{empty}
\vskip 1cm
\centerline{${}^1$\it Sternberg Astronomical Institute, 119899 Moscow,
  Russia} 
\centerline{sorokina@sai.msu.su, oleg@sai.msu.su}
\centerline{${}^2$\it Institute for Theoretical and Experimental Physics,
  117259 Moscow, Russia}
\centerline{sergei.blinnikov@itep.ru, blinn@sai.msu.su}
\centerline{${}^3$\it MPI f\"ur Astrophysik, 85740 Garching, Germany}
 
\vskip 1cm
\centerline{ABSTRACT}
\vskip 0.8cm
 
 The calculations of the light curves
 of thermo\-nuclear supernovae are carried out by a method of multi-group
 radiation hydrodynamics.
 The effects of spectral lines and expansion opacity are taken into account.
 The predictions for {\it UBVI } fluxes are given.
 The values of rise time for {\it B}  and {\it V}  bands found in our 
 calculations are in good agreement with the observed values.
 We explain why our results for the rise time have more solid physical
 justification  than those obtained by other authors.
 It is shown that small variations in the chemical composition
 of the ejecta, produced in the explosions with different 
 regimes of nuclear burning, can influence drastically
 the light curve decline in the {\it B} band and, to a lesser
 extent, in the {\it V} band.
 We argue that recent results on positive cosmological
 constant $\Lambda$, found from  the high redshift supernova observations,
 could be wrong in the case of possible variations of the preferred mode of
 nuclear  burning in the earlier Universe.
\end{titlepage}

\section*{Introduction}\lvm
\label{intr}
 
For decades, Type~Ia supernovae (SNe~Ia) are treated as very suitable objects
for distance measurements in the Universe and for deriving cosmological
parameters (Sandage, Tammann, 1997; Ruiz-Lapuente, 1997).
There are several reasons for this.
The first one is the intrinsic brightness of the SNe~Ia.
Due to their luminosity, one can measure an appreciable flux
even if they are at high redshift, like $z \sim 0.5 - 1$. 
Besides that, the spectra and light curves of
\sn\ seem very uniform at first glance.
However, more thorough investigation demonstrates the diversity within this
class of objects (Pskovskii 1977, Bartunov \& Tsvetkov 1986,
Branch 1987, Barbon et al. 1989, Phillips et al.\
1987; Filippenko, et al.\ 1992, Leibundgut, et al.\ 1993).
 
Quite a while ago, Pskovskii (1977) has shown that there is a correlation 
between
the maximum luminosities of nearby SNe I and their post-maximum decline rates.
This dependence was confirmed subsequently from the observations of many
low-$z$ super\-novae and studied by several workers (Bartunov \& Tsvetkov 1986,
Phillips 1993, Hamuy et al. 1995,1996a, Hamuy \& Pinto 1999).
 
Modern observational technique allows discovering and studying distant,
high-$z$, super\-novae.
The first results were obtained by N\o rgaard-Nielsen et al. (1989):
a few years of their observations have brought up only two super\-novae at
$z \sim 0.3$.
 
Presently, there are several search groups in the world which observe distant
super\-novae with the largest terrestrial telescopes (and sometimes with
the help of the {\it Hubble Space Telescope}).
The observational technique is advanced to the level allowing 
the discovery of dozens
high-$z$ super\-novae per one observational period of a couple of weeks.
The observational material produced by those groups allowed to estimate the
cosmological parameters such as the Hubble constant~$H_0$, the matter density
$\Omega_{\rm_M}$
and the vacuum energy $\Omega_{\Lambda}$ in units of  the critical density
(for definitions see, e.g., Carroll et al. 1992).
The dependent quantities, like the deceleration parameter~$q$, the ratio of
the local value of $H_0$ to the global one, etc., can be derived based on this.
E.g., Kim et al. (1997) have estimated $H_0$ using the first seven \sn\ at
$z \ge 0.35$ and refuted the suspicion that the local value of $H_0$
is appreciably larger than the average one.
Perlmutter et al. (1997) have estimated a probable ratio of the
$\Omega_{\rm M}$ and  $\Omega_{\Lambda}$.
 
Quite recently, several interesting papers appeared based on
a richer statistics of distant super\-novae
(Schmidt et al. 1998;
Garnavich et al. 1998a,b; Riess et al. 1998; Perlmutter et al. 1999).
An extremely important result is claimed in that work: the analysis of the
observations implies with high confidence that the expansion of the Universe
is accelerating at present epoch.
 
One should note, though, that those results on distant supernovae were
based on the maximum luminosity -- decline rate relation derived from the
analysis of nearby objects (in the work by Perlmutter et al. 1999, the high
redshift \sn\ are also used in obtaining such a relation, but this does
not help in excluding possible effects of evolution, see Drell et al. 1999). 
Even for the nearby \sn\ the deviations of
individual objects from the relation cannot be explained solely by observational
errors. The physical understanding of the peak luminosity -- decline rate 
correlation is crucial for estimating the cosmological results obtained
with \sn.
 
In principle, the  slower decline rate for the higher maximum luminosity can
be explained from theoretical point of view:
both are caused mainly by \nifsx\  produced during the explosion.
Radioactive decay of \nifsx\ forms the SN~Ia light curve and determines
its maximum luminosity.
On the other hand,
large amount of nickel should enhance the opacity of matter.
It takes longer time for radiation to diffuse through the stellar matter
and the light curve becomes less steep.
However, the decline rate is influenced not only by the amount of nickel 
(and of other heavy elements), 
but also by their distribution inside the expanding star and
by the velocity distribution of ejected gas.
These distributions are dependent, in turn, on the mode of burning propagation
in the star.
 
After the pioneering work by Arnett (1969), Ivanova et al. (1974),
Nomoto et al. (1976) the theory of thermonuclear burning in supernovae has
been advanced appreciably, yet many problems are still not solved,
see, e.g., Niemeyer and Woosley (1997), Reinecke et al. (1999), Niemeyer
(1999).
A plenty of models for SN~Ia explosions were built, with various masses
(Chandrasekhar and sub-Chandrasekhar), burning modes (detonation,
deflagration, and a variety of their combinations) in a range of explosion
energies and expansion velocities.
As a result of dissimilar burning, chemical elements are produced  in various
proportions and distributed differently throughout the ejecta, even if
conditions in a star prior to the explosion are identical.
This leads to noticeable variations in theoretical light curves.
From the comparison of calculated light curves with the observed ones one can
judge which mode of explosion is preferred in nature.
Different explosion scenarios imply a scatter in the dependencies between
the observable parameters of the burst, and could explain the events deviating
from the standard relations to a larger extent than a margin anticipated from
the observational errors. 

An extensive set of the models for light curves of \sn\ was produced by
H\"oflich et al. (1996,1997), giving
some  theoretical insight into the observed correlations of the peak
luminosity and the post-maximum decline rate. We have undertaken an
independent study of Chandrasekhar and sub-Chandrasekhar SN~Ia light curve
models. The results will be published elsewhere (Sorokina et al. 1999),
some of our calculations for  sub-Chandrasekhar models are discussed by
Ruiz-Lapuente et al. (1999).
In this report we present the results of the calculations of the light curves
for two well-known  SN~Ia
models: the deflagration model W7  (Nomoto et al. 1984)
and the delayed detonation one, DD4 (Woosley \& Weaver 1994).
They both are Chandrasekhar mass models and contain  similar amount of \nifsx,
but the latter is distributed differently due to different burning modes.
Our result is that the two light curves
are not similar in their decline rate, despite the close absolute 
luminosities at maximum light.
The {\it B}-band flux of W7 goes down too slowly and agrees with the
observations of typical \sn\ only marginally.
We explain why the results of other authors somewhat disagree with ours 
for this model.
 
Discussing the possibility of the use of \sn\ in cosmology, we suggest that
the available statistics of distant supernovae does not yet allow drawing firm
conclusions on the geometry of the Universe.
In terrestrial experiments, one cannot predict with certainty the outcome of
an explosion.
The situation for supernovae can be similar:
it is quite likely that initial conditions influence only the probability
of the development of a certain burning mode, but do not determine it
exactly.
Since the light curve shape depends strongly on the regime of burning,
one cannot predict deterministically the decline rate knowing only the initial
conditions.
The probability of the specific decline rate, which is crucial for obtaining
the cosmological parameters, would be found only with a sufficiently rich
statistics of \sn\ at various redshifts.

\section*{Numerical method}\lvm
 
The approach used for light curve modeling
 in our study is the multi-energy group radiation hydrodynamics.
The code we have
used is called {\sc stella} (Blinnikov \& Bartunov
1993; Blinnikov \etal 1998).
It solves the time-dependent equations for the angular
moments of
intensity averaged over fixed frequency bands, using up to $\sim 200$ zones
for the Lagrangean coordinate and up to 100 frequency bins
(i.e., energy groups).  This number of frequency groups allows one
to have a reasonably accurate representation of non-equilibrium continuum
radiation. There is no need to ascribe any temperature to the radiation:
the photon energy distribution can be quite arbitrary.
 
     The  coupling of multi-group radiative transfer with hydrodynamics
means that we can obtain {\it UBVRI} fluxes in a self-consistent
calculation, and that no additional estimates of thermalization depth as in
the one-energy group models  are needed.
Variable Eddington factors are computed, which fully take into account
scattering and redshifts for each frequency group in each mass zone.
The parameters of the decays of \nifsx\ to \cofsx\ and of \cofsx\ to \fefsx\
are taken from the work of Nadyozhin (1994). 
The positron energy is deposited locally.
The gamma-ray transfer is calculated using a one-group approximation for
the non-local deposition of the energy of radioactive nuclei.
Here we follow  Swartz et al. (1995)
and we use purely absorptive opacity in the gamma-ray range.
Swartz et al. addressed this question
using in particular the W7 model.
They found a very weak dependence on the optical depth
and derived the error of only 10\% in heating rate due to the gamma deposition
in the SN~Ia model up to 1200 days after the explosion
for the effective opacity $\kappa_\gamma=0.050Y_e$ ($Y_e$ is the total
number of electrons per baryon).
 
Local Thermodynamic Equilibrium (LTE) for ionization and atomic level
populations is assumed in our modeling.
     In the equation of state, LTE ionization and recombination are taken
into account.  The effect of line opacity is treated as an expansion
opacity according to the prescription of Eastman \& Pinto (1993).

 The main limitation of our current light curve code is the LTE approximation.
To simulate some of the non-LTE effects we used the approximation of the 
absorptive
opacity in spectral lines, following the results of non-LTE computations
by Baron et al. (1996), see the discussion in Blinnikov et al. (1998).
We do not pretend yet that the light curves in this approximation are
reproduced absolutely correctly, especially later than $\sim 2$ months after 
the explosion.
But our experience in the light curve modeling of other types of supernovae
(Blinnikov et al. 1998, Blinnikov 1999, Blinnikov et al. 1999)
shows that our results are very reliable for a month or so after the
maximum light. One can see in Fig.~\ref{tmpr} below, that the photosphere
in {\it B} band for the day 35 in our models is still in the outer layers,
so the bulk of the flux is born in the conditions close to LTE.
What is also important, we are interested here mostly  in relative
changes of the light curves for two classical regimes of burning. 
We believe that the relative difference is reproduced quite 
reliably in our computations.

\section*{Burning modes in thermonuclear supernovae}\lvm
\label{modburn}

Currently, there is no doubt that the light of SNe Ia is produced
in the decays of \nifsx\ to \cofsx\ and then to \fefsx. 
The sufficient
amount of initial \nifsx\ as well as the required energy of the explosion can be
naturally produced during the thermal runaway in
a degenerate carbon-oxygen mixture.
Yet, the regime of the nuclear burning in supernovae
is still a controversial issue.

Arnett (1969) 
was the first to model the supersonic
combustion, i.e. detonation, in supernovae. Later,
Ivanova et al. (1974) 
obtained a sub-sonic flame (deflagration)
propagating in a spontaneous regime with pulsations and switching
subsequently to the detonation.
Nomoto et al. (1976) 
modeled deflagration propagating due to a convective heat transfer with
a parameterized flame speed.
Both detonation and deflagration have their merits and problems
in explaining the supernova phenomenon (see, e.g., the review of Woosley
\& Weaver 1986). 
 
A prompt detonation, born near the center of a star,
would incinerate almost the whole star to ashes consisting
of iron peak elements (Arnett 1969). This is in clear contradiction
with observations which show that
intermediate mass elements are abundant in the ejecta.  Thus at some stage
(if not always) the burning must be sub-sonic.  
Here another problem arises: it is
clear that the combustion
must be much faster than suggested by an analysis of propagation
of a laminar one-dimensional flame (Timmes \& Woosley 1992).
 
From the microscopic point of view the one-dimensional nuclear
flame is a wave described  essentially in the same way
as that formulated by Zeldovich \& Frank-Kamenetsky (1938) in spite of
complications introduced by nuclear kinetics and a very high
conductivity of dense presupernova matter.
It is found that the conductive flame propagates in a presupernova far too 
slowly to explain the supernova outburst correctly:
the flame Mach number is of the order of a few percent and less
(Timmes \& Woosley 1992).

A natural way to accelerate the fuel consumption is the development
of instabilities inherent to the flame front.
As  explained in the classical paper by Landau (1944),
the hydrodynamic instability leads to wrinkling
or roughening of the front surface, i.e., to increasing of its area with
respect to the smooth front, and consequently to acceleration of
the flame propagation. In extreme cases, the instabilities can lead to
a transition from the regime of slow
flame propagation  to the  regime of detonation. The role of the Landau
instability in supernovae was pointed out by Blinnikov et al. (1995) 
and Niemeyer \& Hillebrandt (1995b).
As the instability grows, the front
becomes wrinkled and fractal. The detailed consideration of non-linear
stage of Landau instability and the calculation of the fractal dimension
of the flame front for that case
is given by Blinnikov \& Sasorov (1996).
 
Since the flame propagates in the gravity field and the burned ashes
have lower density than the unburned fuel, the Rayleigh--Taylor (RT)
instability is often considered to be the main process governing
the corrugation of the front
(see M\"uller \& Arnett 1986, Woosley 1990a, Woosley 1990b, 
Livne \& Arnett 1993,
Khokhlov 1993).
The RT instability gives birth to a turbulent cascade providing for an
acceleration to the flame and additional difficulties in modeling the SN
event (Niemeyer \& Hillebrandt 1995a, Niemeyer 1995, Niemeyer \& Woosley 1997).
The turbulent mixing of hot products of burning with the cold fuel
might trigger a detonation after a period of slow expansion
of the partly burned degenerate star (Niemeyer \& Woosley 1997). 
Here the transition to the detonation may happen via a stage of supersonic
spontaneous flame propagation (Blinnikov \& Khokhlov 1986).
However, recent computations by Reinecke et al. (1999) show that the RT
instability alone is not sufficient for the explosion: it drives a flame front
too weakly and the needed subsonic speed is not achieved. 
Additional sources for the turbulence or alternative routes for accelerating
flames seem to be needed (Niemeyer 1999). Even if detonation is established,
it cannot be described by a classical Zeldovich--von Neumann wave, which is
unstable in the conditions of \sn\ explosions (see Imshennik et al. 1999 and
references therein).

It is clear that a full direct modeling of the burning is not possible because
the involved scales range from the thickness of a flame $ \sim 10^{-5}$ cm 
up to the dimension of a white dwarf,  $\sim 10^9$~cm.
Moreover, in all cases where turbulence
is involved, the chaotic behavior sets in inevitably, and a subtle difference
in initial conditions can lead to drastically unsimilar outcomes. This
is well known from terrestrial experiments: after a series of tens or
even hundreds cases of slow burning, sometime a detonation sets in
when all the conditions of the experiment are quite the same
(e.g., Mader 1979).

\section*{Theoretical Models of Supernovae and Light Curves}\lvm
 
\begin{figure}
\plotone{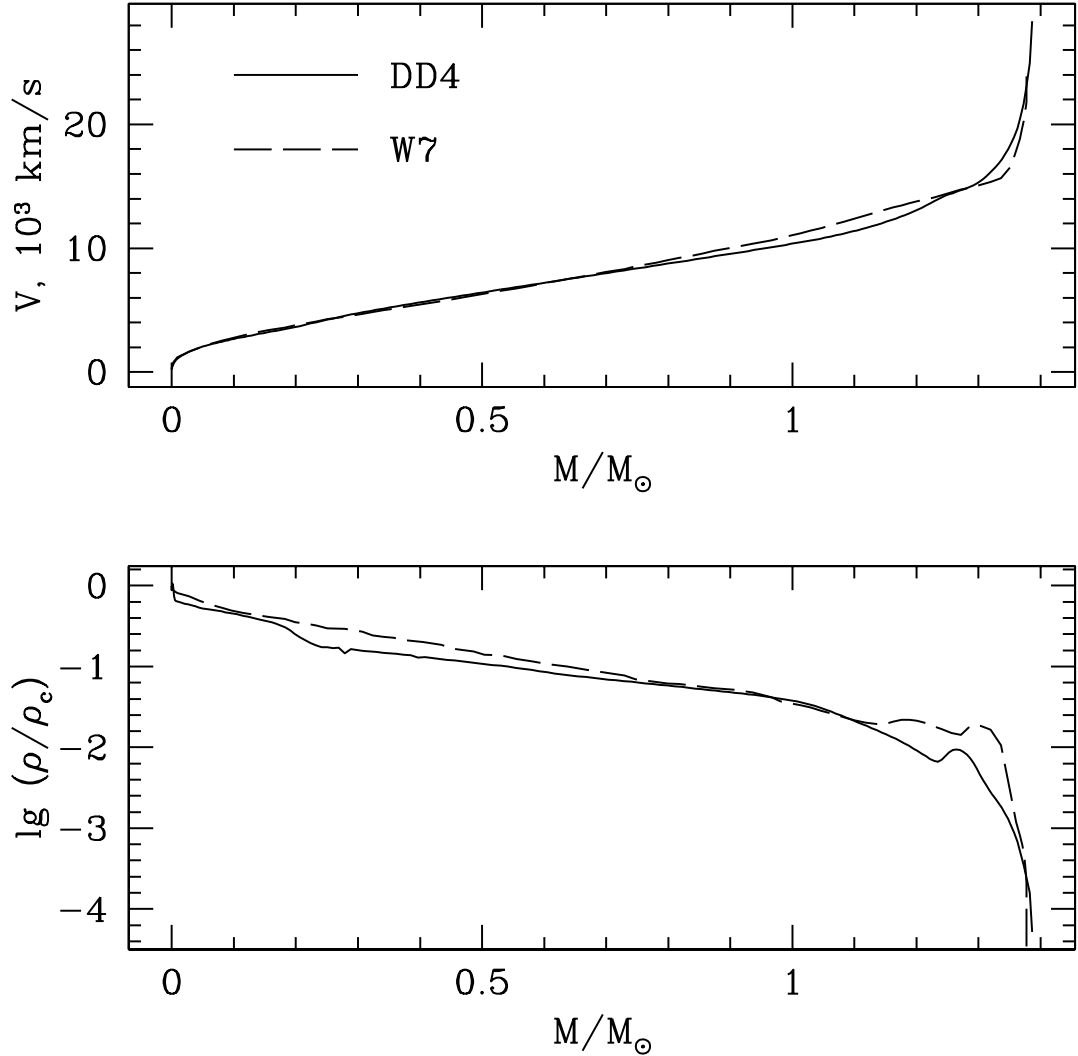}
\caption{The density and the velocity distributions on mass at the beginning of
our calculations of models DD4 and W7. The density is plotted relative to
the central value. We have started 
from $\rho_c = 9.76\cdot10^{-9} {\rm g~cm}^{-3}$
at $t = 8.16\cdot 10^4$~s for DD4, and from 
$\rho_c = 7.42\cdot10^{-3} {\rm g~cm}^{-3}$ at $t = 8   \cdot 10^2$~s for W7.}
\label{rhov}
\end{figure}

\begin{figure}
\plotone{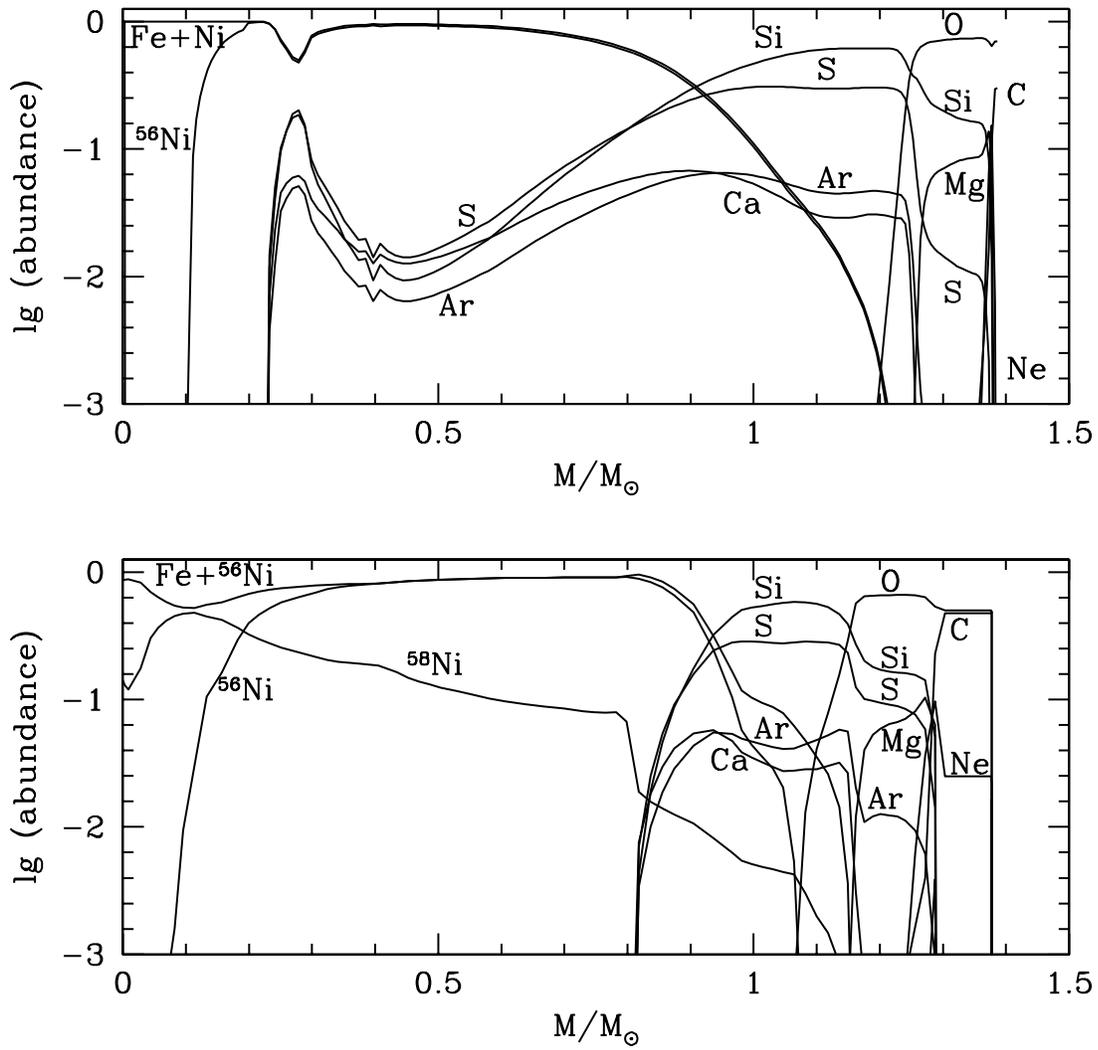}
\caption{The distribution of the most abundant chemical elements 
throughout the ejecta for DD4 (top) and W7 (bottom).}
\label{difchem}
\end{figure}

Initial models corresponding to two different regimes of burning were used in
our calculations, namely, the classical deflagration model W7
(Nomoto et al. 1984)
and a delayed detonation model DD4 (Woosley \& Weaver 1994).
The main parameters of these two models are compared in Table~\ref{modpar},
and their resulting chemical compositions, in Table~\ref{modchem}.
Figures~\ref{rhov}, \ref{difchem} display the distributions of
density, velocity, and basic chemical elements in the ejecta for those two 
models.
It is clear from the figures and tables that the hydrodynamical parameters are
approximately the same for both models, while differences in chemistry are due
to the relevant burning mode in a white dwarf (though, the amount of \nifsx\
is almost the same in both models, so maxima in luminosity are close).
Both W7 and DD4 are simple 1D models, and each individual SN~Ia can be
a more complicated 3D fractal event in reality.
Yet our computed light curves have a good resemblance
to the observed ones and reproduce the broad band fluxes quite well.
 
\begin{figure}
\plotone{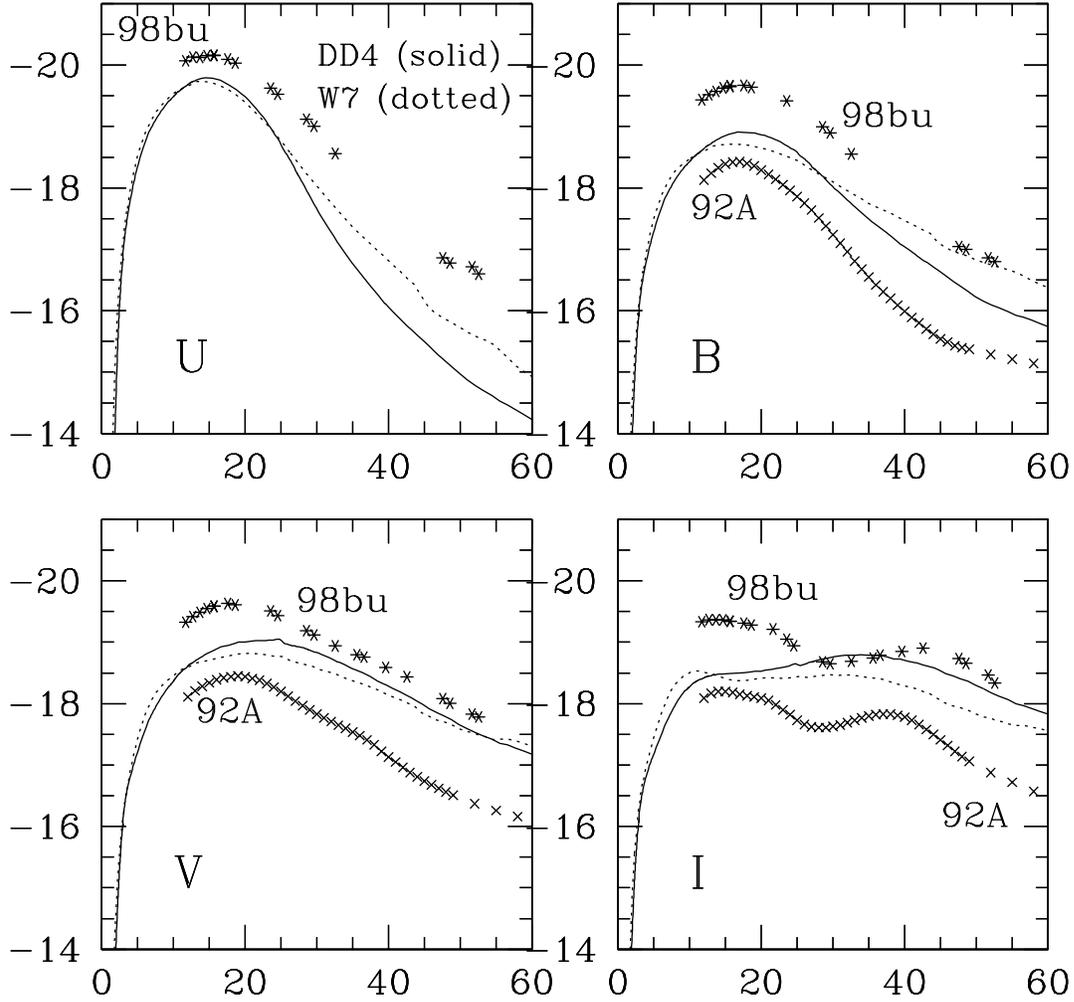}
\caption{The simulated {\it UBVI } light curves of DD4 (solid line) and W7
(dotted line). Asterisks and crosses correspond to the light curves of the
SN~1998bu and SN~1992A, respectively. The observed light curves are shifted
along the time axis so that their maxima in {\it B} occur at the same time
as the maximum of the calculated flux from DD4.}
\label{w7dd4}
\end{figure}

The curves in the Fig.~\ref{w7dd4} display the numerical results 
for the models W7 and DD4, while asterisks show the observational points 
for  SN 1998bu
(Suntzeff et al. 1999; see also Meikle \& Hernandez 1999, and Jha et al.
1999) 
and crosses the template fluxes for SN 1992A (Hamuy et al. 1996b).
The results demonstrate clearly that in spite of similar maximum luminosities,
the models have very unequal decline rates in the {\it B} filter,
though they are much more similar in the {\it V} band.
 
It is not easy to explain this scatter in the predicted decline rates by a 
single factor.
The outgoing flux of the SN~Ia explosion is influenced by many physical
parameters.
The differences in all parameters for the models are small for each instant.
Yet the combination of the minute differences seems to be sufficient
to change the light curve shapes.
For example, the distributions of \nifsx\ in outer layers, as well as lighter
elements closer to the center, are not the same, as shown in the
Fig.~\ref{difchem}.
This causes slight changes in gamma-ray deposition and in diffusion of soft
photons, thus implying unsimilar temperature distributions.
The latter are shown in Fig.~\ref{tmpr} for the day 35 after the explosion.
The solid dots mark the layers where the optical depth becomes equal to
$2/3$, if measured from the outer boundary of the ejecta.
The  temperature for W7 is 10\% higher, and the radius is also a bit larger
than for DD4 in this point.
This can explain quite well the differences in {\it B} fluxes for this epoch
(see Fig.~\ref{w7dd4}).
 
Let us compare  our computed light curves with observations of
real supernovae.
There are several quantitative relations to describe their shapes.
One of the most popular is the dependence suggested by Phillips (1993),
$M_{\rm max}(\Delta m_{15}^B)$,
where $\Delta m_{15}^B$
is a drop of the {\it B} flux (in stellar magnitudes) 15 days
after the maximum.
The linear regression fit
$$M_{\rm max} = a + b \Delta m_{15}^B$$
is usually used for the interpretation of observations.
The values of $a$ and $b$ vary strongly, when obtained with different
sets of \sn\ by different authors.
Moreover, the dispersion of their values is large even within one set.
We are more interested in the value of $b$, the slope of the light curve.
Its variation is rather large,
e.g., Phillips (1993) obtains $b^B = 2.698\pm 0.359$ from the observations of 9
nearby supernovae.
In contrast, analysis by Hamuy et al. (1995) of only six supernovae from the
same set yields $b^B = 1.624\pm 0.582$.
This scatter cannot be explained only by the observational errors,
they are not so big.
See, however, Hamuy et al. (1996a) where they obtain much smaller scatter
and even lower value of $b^B = 0.784\pm 0.182$ using 26 Cal\'an/Tololo
distant supernovae.
The value of $a$ is influenced by the Hubble parameter~$H_0$:
$a=a_0+5\lg(H_0/85)$ for Phillips (1993), and
$a=a_0+5\lg(H_0/65)$ for Hamuy et al. (1996a).
In reality, supernovae do not lie on a straight line, they occupy an extended
region on the diagram $M_{\rm max}(\Delta m_{15}^B)$.
Most probably, this reflects real variations of the explosion physics.
One can come to the same conclusion after analyzing the correlations between
other key parameters of SN~Ia, like \nifsx\ mass, peak bolometric
luminosity, rise time to the maximum light, explosion energy, etc
(Nadyozhin, 1997).
 
The slope  of our calculated light curves (especially
for DD4 model) and the luminosity are comparable with the observable values.
We can estimate the decline rate from the results of Phillips (1993) and
Hamuy et al. (1995) with the formula
$\Delta m_{15}^B = (M_{\rm max}^B - a)/b$.
Comparing this with the rates for W7 and DD4 in our computations we find
for $H_0=85$ km/s/Mpc:
$$
 \Delta m_{15}^B = \left\{
    \begin{array}{llll}
        1.04(\mbox{Phillips}) & 0.94(\mbox{Hamuy95})
               & 1.07 (\mbox{model DD4}) & \quad
                                \mbox{for } M_{\rm max}^B = -18.9;\\
         1.12(\mbox{Phillips}) & 1.06(\mbox{Hamuy95})
               & 0.66 (\mbox{model W7}) & \quad
                                \mbox{for } M_{\rm max}^B = -18.7.\\
    \end{array}
                         \right.
$$
So, we see that the model W7 behaves in the sense opposite to the 
Pskovskii-Phillips relation: being a bit weaker than DD4 it declines slower,
instead of fading down faster. 

If we take more recent data from Hamuy et al. (1996a) and
$H_0=65$ km/s/Mpc, then we get higher values for $ \Delta m_{15}^B$,
predicted by `inverse Phillips' relations,
$\Delta m_{15}^B = (M_{\rm max}^B - a)/b$,
for the $ M_{\rm max}^B$  found in our
modeling, but this
reflects only the fact that both models are underluminous for the long
cosmological distance scale. We are interested in relative differences between
two models, so it is more instructive to compare the absolute magnitudes
from the direct formula
$M_{\rm max} = a + b \Delta m_{15}^B$. We present the comparison for the two
values of the Hubble parameter.
For $H_0=85$ km/s/Mpc:
$$
 M_{\rm max}^B  = \left\{
    \begin{array}{llll}
        -18.8(\mbox{Phillips}) & -18.7(\mbox{Hamuy96})
               & -18.9 (\mbox{model DD4}) & \quad
                                \mbox{for }\Delta m_{15}^B = 1.07;\\
         -19.9(\mbox{Phillips}) & -19.0(\mbox{Hamuy96})
               & -18.7(\mbox{model W7}) & \quad
                                \mbox{for }\Delta m_{15}^B  = 0.66 \\
    \end{array}
                         \right.
$$
Thus, for the short distance scale DD4 is OK, while W7 will look underluminous.
For $H_0=65$ km/s/Mpc:
$$
 M_{\rm max}^B  = \left\{
    \begin{array}{llll}
        -19.4(\mbox{Phillips}) & -19.3(\mbox{Hamuy96})
               & -18.9 (\mbox{model DD4}) & \quad
                                \mbox{for }\Delta m_{15}^B = 1.07;\\
         -20.5(\mbox{Phillips}) & -19.6(\mbox{Hamuy96})
               & -18.7(\mbox{model W7}) & \quad
                                \mbox{for }\Delta m_{15}^B  = 0.66 \\
    \end{array}
                         \right.
$$
Now, both models are too weak, and this means that for the   long
distance scale a standard model, like DD4, which is 0.4 magnitudes 
underluminous according to the peak luminosity--decline rate relation
of Hamuy et al. (1996a), must have perhaps 30 to 40\% more
\nifsx\ (Arnett 1982, Cappellaro et al. 1997, Mazzali et al. 1998).
At the same time W7 is underluminous already by 0.9 magnitudes.
If we assume the same relative difference between delayed detonation
and deflagration models with the same enhancement of \nifsx\ abundance, 
then the W7-like events will still look 0.5 magnitudes underluminous. 
This can have drastic implications for the use of \sn\ in cosmology.

One can see in the Fig.~\ref{w7dd4} that the  behavior of both
calculated light curves is qualitatively  correct when compared with
a couple of well studied individual \sn.
In particular, the rise time corresponds to the observed values:
the flux of observed \sn\ grows up to the maximum during 15--20 days
(Maza 1981, Pskovskii 1984, Bartunov \& Tsvetkov 1997).
As it is shown in Table~\ref{rise}, our calculations predict 16--17 days 
for this value in the $B$ band,
and this is in much better agreement with the observations than yielded by
previous computations of
H\"oflich et al. (1997).
(Their fluxes grow for only 10--15 days for the majority of models, including
W7.)
One can also note the realistic behavior of our computed light curves in
different spectral bands.
The decline rate is less steep for redder bands in both models (see
a simple analytical explanation for this given by Arnett 1982)
But it is also evident that the deflagration model fades down 
in the {\it B}-band too slowly, and
does not fit the decline rates for typical type Ia supernovae,
while the delayed detonation model agrees with the observations very well.
 
Until recently W7 has been in frequent use in many papers where light curves
of \sn\ were presented,
as well as when supernova remnants were calculated  subsequently.
The reason for such a popularity is that this model seemed to reproduce the
observed fluxes quite well (see, e.g., H\"oflich 1995 and Eastman 1997).
Our results for W7 (especially in the {\it B} band) differ from the ones
obtained in the previously published papers.
This can  be explained most probably by the differences in including
the expansion opacity (Karp et al. 1977) into the energy equation.
The correct treatment of the energy equation is described by
Blinnikov (1996, 1997) who
has shown that there is no effect of expansion opacity 
when line extinction is taken into account in the energy
equation for photons. The expansion effect is important only in the flux 
equation.
Extinction in lines is much stronger than other sources of
opacity for the \sn.
So when the effect by Karp et al. is included into the energy equation,
this leads to unjustifiable enhancement of the energy exchange between 
radiation and matter.
The photon energy is in this case artificially redistributed towards longer
wavelengths, and escapes the system much easier.
As a result, the flux from a supernova grows up and falls down faster,
and the slope of the light curve becomes steeper.
 
\begin{figure}
\plotone{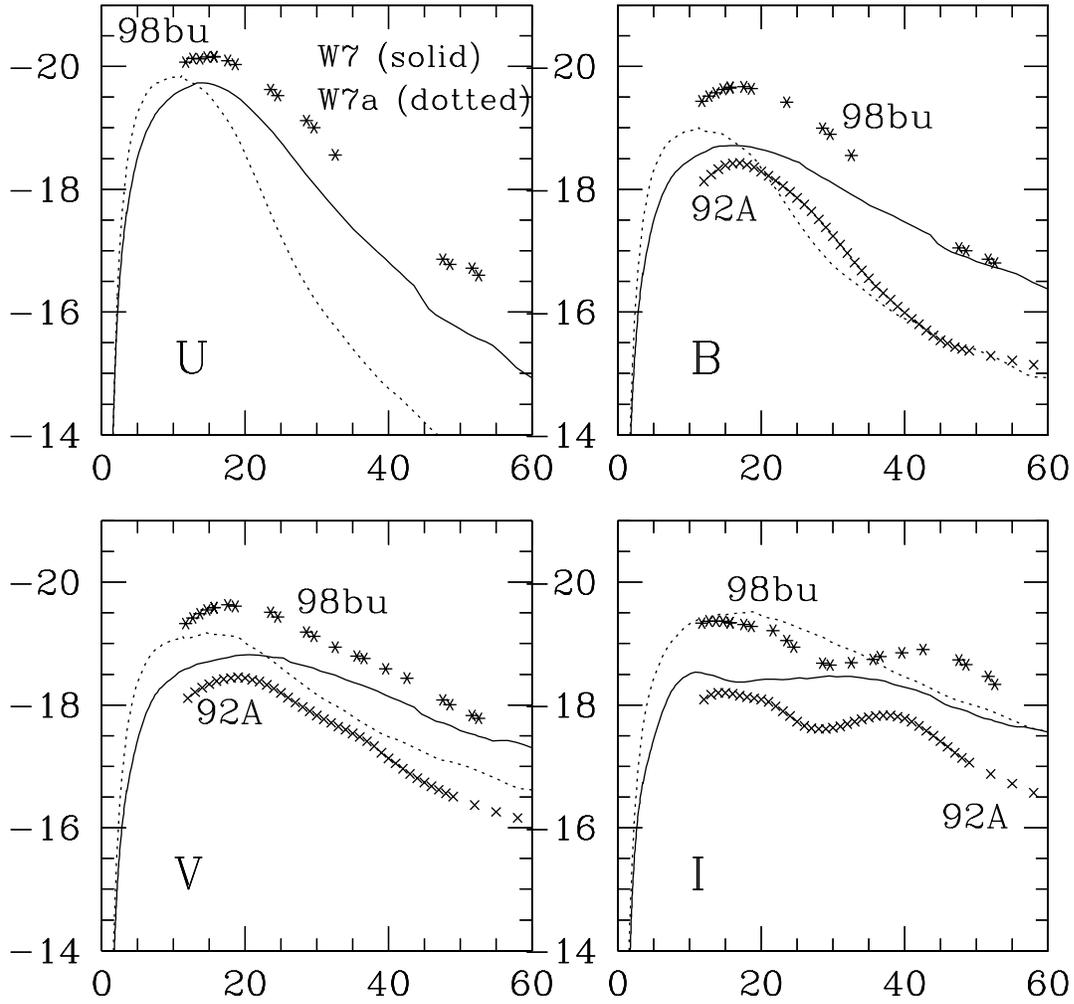}
\caption{The comparison of the light curves for two runs of the model W7.
\hspace{1mm} 
a)~The extinction is treated as pure absorption and thereby the expansion line
opacity is fully (and incorrectly) included into the  energy equation 
(the model W7a, dotted). \hspace{1mm}
b)~Scattering is taken into account; the expansion effect  in line opacity 
is not
included in the energy equation, i.e. the correct equation is solved
(the model W7, solid.)}
\label{abscat}
\end{figure}

To check the importance of this effect,
we have calculated the model W7 once more,
but this time we treated the extinction (including the one in lines and the
expansion effect) as a pure absorption.
In such a way, we have made the energy exchange stronger, and so we have
spoiled the energy equation.
Because of this, the decline rate of the light curve in the {\it B} band has 
become ``better'', although the rise time is shortened and its
agreement with the observations is much worse.
One can notice here that the results of the calculations of our ``spoiled''
W7 model resemble those by H\"oflich.
Fig.~\ref{abscat} shows how the enhanced energy exchange influences 
the light curve of W7.
It is easy to see from the figure that although the flux in shorter
wavelengths could be called ``good'' (but with shortened rise time; see also
Table~\ref{rise}), 
it would be absolutely untrue regarding the flux in {\it I}.
The secondary maximum in {\it I} band, which is observed in almost all real
\sn, has disappeared in the new light curve.
Physical processes responsible for the presence of this feature in the
light curves should yet be investigated.
Nevertheless, the lack of the secondary maximum clearly demonstrates
that some important processes were not included (or were included
incorrectly) in the treatment of radiative transport.
 
Thus, we are able to reproduce the results of other researchers 
when the wrong energy equation is used in our code.
When the appropriate physics is included instead, the decline rate
of the deflagration model W7 is too slow, and does not
correspond to the behavior of a typical SN~Ia.
Two possible reasons for such a result could be proposed:
either the real explosions are not due to the pure deflagration or, most
probably, too many simplifications were supposed while constructing W7.
More complicated deflagration model could, perhaps, describe the reality
better (cf. Niemeyer \& Woosley 1997, Reinecke et al. 1999).

\section*{Using SNe Ia in Cosmology}\lvm

Due to the maximum light -- decline rate dependence, 
\sn\ are now in active use for finding cosmological distances. 
Their observations can help to determine the geometry of the Universe, as
well as to check the viability of some cosmological models.
The idea is to compare the photometric distance to a supernova with its
value in a given cosmological model when its redshift $z$ is measured.
The simple formula $d_{\rm ph}=\left(L/4\pi F\right)^{1/2}$ defines the
photometric distance to an observed object if its absolute luminosity $L$ is 
known.
At the same time, the photometric distance can be calculated theoretically:
if $z$ is fixed, then $d_{\rm th}$ is a function of the cosmological parameters $H_0$,
$\Omega_{\rm M}$ and $\Omega_\Lambda$ (Carroll et al. 1992):
$$
d_{\rm th}={c\over H_0}(1+z){1\over\sqrt{\Omega_k}}\,\sinh\left\{
          \sqrt{\Omega_k}\int\limits_0^z \left[
             \Omega_{\rm M} (1+z)^3  +\Omega_\Lambda
            +\Omega_k (1+z)^2
                                         \right]^{-1/2} {\rm d}z
                                                         \right\}.
$$
Here $\Omega_k=1-\Omega_{\rm M}-\Omega_\Lambda$,
and for  $\Omega_k<0$ the hyperbolic sine ($\sinh$) transforms, according to
elementary formulae, to  the trigonometric sine ($\sin$), while
$\sqrt{\Omega_k}$ should read $\sqrt{|\Omega_k|}$. For $\Omega_k=0$ the limit
$\Omega_k \rightarrow 0$ is easily taken, so  $\sinh$ disappears
from the expression for $d_{\rm th}$, and only the integral is left. 
 
At present, several search groups in the world do observe the high-$z$ supernovae.
The peak luminosity -- decline rate relation is used by them to estimate the
cosmological parameters from these observations.
The relation is calibrated with the help of nearby supernovae samples,
since rich data are available in this case (see, e.g., Phillips 1993),
but it is used to interpret the observations of the much more distant 
supernovae as well. 
We already discussed that even nearby examples are not fully homogeneous,
though they seem to be equivalent at first glance.
The investigations of the influence of the evolutionary effects 
on the light curve shape are only at the beginning so far.
 
The maximum light -- decline rate relation could be different for
younger galaxies. 
To understand how it could change, one should know how the shape of the
light curve depends on the initial chemical composition of the exploded white
dwarf. And what are the initial conditions in general, what are the
progenitors of supernovae, do those conditions change with the age of
Universe?
 
The influence of the initial conditions just before the explosion on the
emission of the \sn\ was studied by Dom\'\i nguez et al. (1999).
 The
role of evolution of \sn\ is discussed 
by Riess et al. (1998), Schmidt et al. (1998) and Drell et al. (1999).
(See also the discussion on different pre-supernova models 
in von~Hippel et al. 1997, Livio 1999, and Umeda et al.\ 1999.)
 
Schmidt et al. (1998) compare the distances to 8 SNe Ia in early type galaxies
with 19 in late type ones and find no significant difference. But here are
two points that need be yet clarified.  First, it is hard to select
the progenitors of the same ages in the late type hosts. Second, 
the statistics of such progenitors is too small. 
One has to build distributions
of SNe Ia at different $z$ with good statistics and to compare them.
The statistical analysis by Drell et al. (1999) shows convincingly 
that it is virtually impossible to distinguish the
effects of evolution from the effects of the geometry of the Universe
expressed through $\Omega_{\rm M}$ and $\Omega_\Lambda$.  Drell et al.
(1999) write that without physical understanding, the additional data, 
such as source spectra, cannot be
used to argue {\it against} evolution, while observers
pretend  that there is no compelling evidence
{\it for} evolution  in the SNe Ia samples, based on these data.
The physical insight in the problem is being developed, e.g. by 
H\"oflich, Wheeler, and Thielemann 1998, Dom\'\i nguez et al.\ 1999,
Umeda et al.\ 1999.

Dom\'\i nguez et al. (1999) find that due to the changes in metallicity and
C/O ratio, the light curve shape can also change.
This  leads
to the drift of $\sim 0.2^{\rm m}$ in distance modulus determined with the
{\it averaged} luminosity -- light curve shape relation.
This effect was considered for only one set of delayed detonation models,
and the results were found to be influenced
mostly via the change of \nifsx\ mass.
 
All previous work on the evolution of the SN~Ia progenitors is based 
on the assumption that chemistry and
hydrodynamic structure of the pre-supernovae exactly fix the behavior of the
light curve.
This is a deterministic
approach. With this approach the net effect is not necessarily large.
As we find in this work, the supernova emission may also depend drastically 
on the mode
of burning, and this is most probably  the factor that cannot be predicted
in a deterministic way in principle.
The main difference in our approach is a conjecture that even for the same 
initial conditions (or minute differences in the initial conditions) 
the outcome of the explosion can be vastly different. 
For the same chemistry and for
the same age of progenitors there can be a wide distribution of different  
SN Ia events, and this is confirmed by the observations 
(see, e.g, Hamuy et al. 1995, 1996a). 
If this distribution (there is no sense to speak
about individual events!) changes with $z$ then one can obtain a wrong
value of $\Lambda$.
Fig.~\ref{w7dd4} is a bright illustration of the possibility to erroneously
determine the absolute flux from a SN~Ia and, therefore, its distance.
It shows the strong deviation in the decline rate among the \sn\ with 
similar maximum light due to the differences in the burning regime, 
which is hardly predictable in principle, as was discussed above. 
One can only predict a trend for the preferred mode of burning, which can 
indeed depend on the age and the chemistry of the progenitor.
 
Let us assume that W7 and DD4 represent two limiting cases   
for real Chandrasekhar-mass explosions. There can be a continuous set
of supernovae between those limits, or there can be a clustering to the
two extreme cases. This does not matter for our purpose now. 
What matters is the indeterminacy principle governing the behavior 
of the flame inside a star. 
Thus, when the initial conditions are fixed, there can be a trend 
in the distribution of the real events to be more ``DD4-like'' or more
``W7-like''.
If, for instance, we have more W7-like events at high $z$ then we can obtain 
a wrong result when we try to derive the distance, and hence, 
the cosmological constant from SNe Ia, based on the local calibration
samples, because of different luminosity -- slope
dependence in the past.

In the recent work on the cosmological supernovae (Riess et al. 1998,
Perlmutter et al. 1999) the
conclusion is reached, that the distance to the high-$z$ supernovae is about
10--15\% larger than expected for the model with $\Omega_{\rm M}=0.2$,
$\Omega_\Lambda=0$.
It is claimed that a model with $\Omega_\Lambda>0$ is needed to avoid such a
deviation.
In reality, $\Lambda$ could be exactly  zero if the fraction of slowly 
burning supernovae would be larger for higher $z$. 
The true luminosity would be then smaller
than found from the peak luminosity -- light curve shape calibrations
for the local
sample, and this could mimic the effect of positive $\Lambda$.
We are persuaded that so far there is no sufficient statistics of distant
supernovae to make firm conclusions on the geometry of the Universe.
Not only the non-zero $\Lambda$ can reconcile the theoretical distances 
$d_{\rm th}$ with
the observed photometric ones,
but also the evolution of the probability of the burning regimes,
making the W7-like explosions in the present epoch less likely than in
the younger Universe.

\section*{Conclusions}\lvm
 
We have chosen as initial configurations for \sn\ 
two models of carbon-oxygen white 
dwarfs with different regimes of burning during the explosion, and simulated
the light curves till $\sim 2$ months after the burst.
We find that the post-maximum decline rate is very sensitive to the
burning mode.
Even minute differences in the \nifsx\ distribution seem to be sufficient to
heat the material differently inside the ejecta and  to
produce different slopes of the light curve thereby, despite the similar 
maximum brightness.

The combustion theory on its modern level is not able to predict the mode of
burning if the initial conditions are fixed. It is conceivable that this
prediction is impossible in principle. 
The different burning modes, in turn, can result in unlike temperature
distributions.
If this is true, then the initial conditions do not  determine the absolute 
maximum -- decline rate dependence exactly.
They fix only the probability to obtain one or another slope of the light
curve for the fixed maximum brightness.
This probability can be changed by variation in the initial conditions.
 
One can speculate that the future combustion theory will be able 
to predict only a proportion between ``fast'' and ``slow'' supernovae,
and this proportion can be very sensitive to the initial conditions.
Astronomers should not wait for such an advancement of the theory,
but could already now check the distribution function of maximum brightness 
over the decline rate at different $z$.
 
What to do in this situation? First, the statistics must be improved to such
an extent that, at a given high redshift,
one could compare the distributions of SNe 
on the maximum luminosity -- decline rate diagram. 
I.e., at each $z$ one should build the distributions of the decline rates
in a broad range of the peak luminosity. The hardest thing is, of course,
to observe the intrinsically weak \sn\ for sufficiently long time.
The distributions found for high  $z$ could coincide with the one for nearby 
supernovae, but could also differ from it.
The latter will mean that nearby statistical relations are not applicable 
to the distant events.
But even in the case when the brightest supernovae at, 
for instance, $z = 1$
have the same distribution of their light curve slopes as the brightest 
samples at $z < 0.1$,
this will tell no word about the coincidence of their {\it absolute}
magnitudes if one does not use any additional data.

Second, the individual differences are larger in the {\it I} band.
This spectrum interval is more sensitive to the explosion physics.
Perhaps,  
the influence of the burning mode on the {\it I} band will become more
transparent in future theoretical investigations.
The theorists should simulate the light curves in {\it I} for extended range 
of models and compare them with the observations.
It is hard
to measure the rest frame {\it I} flux for large $z$, 
since then it goes to the far infrared on Earth. 
But this must be possible with new orbiting telescopes,
like the {\it Next Generation Space Telescope} (Dahl\'en \& Fransson 1998,
Rees 1998).
The observations in {\it I} should help to put useful constraints
on the supernova models for each real explosion.
But the thorough study is still needed both to understand the role 
of the non-LTE effects in the emission of \sn\ in near infrared and 
to develop the realistic 3D models of thermonuclear burning. Ultimately,
the radiative transfer computations for those 3D-models will show the level
of accuracy of 1D models considered here.

\vskip 1cm

We are grateful to P.Ruiz-Lapuente, S.Woosley, K.Nomoto, K.Iwamoto for
providing us with the models used in the calculations,
to P.Lundqvist for the possibility of doing calculations at
Stockholm Observatory and for his warm hospitality during our stay in Sweden.
We are also thankful to D.Tsvetkov for very useful comments.
The work was supported in part by INTAS--RFBR project 95-0832
`Thermonuclear Supernovae', by ISTC grant 370--97, by the Royal Swedish
Academy, and by the University of Barcelona.

\bigskip
 
\centerline{\bf\Large References}
 
 
\begin{description}
 
\item Arnett W.D. 1969, Astrophys. Space Sci.,  5, 180

\item Arnett W.D. 1982, ApJ, 253, 785 

\item Baron~E., Hauschildt~P.H., Nugent~P., \& Branch~D.  1996, 
    MNRAS,  283, 297
 
\item Bartunov~O.S. \& Tsvetkov~D.Yu. 1986, 
	Astrophys. Space Sci.,  122, 343
 
\item Bartunov~O.S. \& Tsvetkov~D.Yu. 1997, 
        in Thermonuclear Supernovae (ed. Ruiz-Lapuente~P. et al.), Dordrecht:
        Kluwer Acad. Pub., p.87
 
\item Blinnikov~S.I. 1996, 
  Astron. Letters    22,  79
  (translated from Russian: Pisma Astron. Zh., 1996,  22, 92)
 
\item Blinnikov~S.I. 1997, in Thermonuclear Supernovae (ed.
	Ruiz-Lapuente~P. et al.), Dordrecht:
	Kluwer Acad. Pub., p.589
 
\item Blinnikov~S.I. 1999, Astron. Letters, in press
  (translated from Russian: Pisma Astron. Zh., 1999)
 
\item Blinnikov S.I. \& Bartunov O.S. 1993, 
   A\&A,  273, 106
 
\item Blinnikov~S.I., Eastman~R., Bartunov~O.S., Popolitov~V.A., \&
 Woosley~S.E. 1998, 
         ApJ,  496, 454
 
\item Blinnikov~S.I. \& Khokhlov~A.M. 1986, Sov.  Astron. Lett.,  12, 131
	(translated from Russian: Pisma Astron. Zh., 1986,  12, 318)
 
\item Blinnikov~S.I., Lundqvist~P., Bartunov~O.S., Nomoto~K., \& Iwamoto~K. 
	1999, 
         ApJ, in press
 
\item Blinnikov~S.I. \& Sasorov~P.V. 1996, 
	Phys. Rev.  E,  53, 4827
 
\item Blinnikov~S.I., Sasorov~P.V., \& Woosley~S.E. 1995, 
	Space Science Rev.,  74, 299

\item Branch, D., 1987 ApJ, 316, 81L
 
\item Barbon, R.,   Rosino,  L. and Iijima, T. 1989, A\&A, 220, 83

\item  Cappellaro E., Mazzali P.A., Benetti S., Danziger I.J., Turatto~M., 
       Della Valle~M., \& Patat~F. 1997, A\&A, 328, 203 

\item Carroll~S.M., Press~W.H., \& Turner~E.L. 1992, 
	Ann. Rev.  Astron. Astrophys.,  30, 499
 
\item Dahl\'en~T. \& Fransson~C. 1998, in The Next Generation
	Space Telescope: Science Drivers and Technological Challenges, 
        (astro-ph/9809379)
 
\item Dom\'\i nguez~I., H\"oflich~P., Straniero~O., \&
	Wheeler~C. 1999,
in Future Directions of Supernovae
Research: Progenitors to Remnants. Assergi (Italy) Sep-Oct 1998
	(astro-ph/9905047)
 
\item Drell~P.S., Loredo~T.J., \& Wasserman~I. 1999, ApJ, in press
	(astro-ph/9905027)
 
\item Eastman~R.G. 1997, in Thermonuclear Supernovae
	(ed. P.Ruiz-Lapuente et al.), Dordrecht:
	Kluwer Acad. Pub., p.571
 
\item Eastman R.G. \& Pinto~P.A. 1993, 
             ApJ,   412,  731
 
\item Filippenko~A.V., et al. 1992, ApJ, 384, L15

\item Garnavich~P.M. et al. 1998a, ApJ,  493, L53
 
\item Garnavich~P.M. et al. 1998b, ApJ,  509, 74
 
\item Hamuy~M., Phillips~M.M., Maza~J., Suntzeff~N.B., 	Schommer~R.A., \&
	Aviles~R.  1995, AJ,  109, 1

\item  Hamuy M., Phillips M.M., Suntzeff N.B., Schommer R.A., 
       Maza J., \& Aviles R. 1996a,  AJ, 112, 2391 

\item Hamuy M., Phillips M.M., Suntzeff N.B., Schommer~R.A., Maza~J., 
Smith~R.C., Lira~P., \& Aviles~R. 1996b, AJ, 112, 2438 

\item Hamuy M.  \& Pinto P.A. 1999,  AJ, 117, 1185 
 
\item von~Hippel~T., Bothun~G.D., \& Schommer~R.A. 1997, AJ,  114, 1154
 
\item H\"oflich P. 1995, ApJ,  443, 89
 
\item H\"oflich~P., Khokhlov~A., Wheeler~J.C.,
      Phillips~ M.M., Suntzeff~N.B., \&   Hamuy~M. 1996,  ApJ, 472, L81

\item H\"oflich~P., Khokhlov~A., Wheeler~J.C., Nomoto~K., \&
	Thielemann~F.K. 1997, in Thermonuclear Supernovae
        (ed. Ruiz-Lapuente~P. et al.), Dordrecht:
	Kluwer Academic Publishers, p.659
 
\item H\"oflich P.,  Wheeler J.C. \& Thielemann F.K. 1998, ApJ, 495, 617

\item Ivanova~L.N., Imshennik~V.S., \& Chechetkin~V.M. 1974, 
	 Astrophys. Space Sci.,  31, 497

\item Imshennik~V.S., Kal'yanova~N.L., Koldoba~A.V., \& Chechetkin~V.M. 1999, 
      Astronomy Letters, 25, 206

\item Jha~S. et al. 1999, to appear in ApJS (astro-ph/9906220)

\item Karp~A.H., Lasher~G., Chan~K.L., \& Salpeter~E.E. 1977, 
   ApJ,   214,  161.
 
\item Khokhlov~A.M. 1993,  ApJ,  419, L77.
 
\item Kim~A.G. et al. 1997, ApJ, 476, L63
 
\item Landau~L.D. 1944, ZhETF,  14, 240; Acta Physicochim. USSR,
        1944,  19, 77 
 
\item Leibundgut B. et al. 1993, AJ, 105, 301

\item Livio~M. 1999, astro-ph/9903264
 
\item Livne~E. \& Arnett~W.D. 1993, ApJ, 
	 415, L107
 
\item Mader C.L. 1979, Numerical Modeling of detonations, Bercley: Univ.
of California Press
 
\item Maza~J. 1981, IAU Circ. 3583

\item Mazzali P.A., Cappellaro E., Danziger I.J., Turatto~M., \& Benetti~S. 
1998, ApJ 499, L49 

\item Meikle~P. \& Hernandez~M. 1999, to appear  in the Journal of the
      Italian Astronomical Society (astro-ph/9902056)

\item M\"uller~E. \& Arnett~W.D. 1986,  ApJ,
	 307, 619

\item Nadyozhin D.K. 1994, ApJS, 92, 527 

\item Nadyozhin D.K. 1997, colloquium ``Supernovae and Cosmology'', Basel
 
\item Niemeyer~J.C. 1995, Ph.D. Thesis, Garching: MPA-911
 
\item Niemeyer~J.C., 1999,  ApJL in press (astro-ph/9906142)

\item Niemeyer~J.C. \& Hillebrandt~W. 1995a, 
	ApJ,  452, 769
 
\item Niemeyer~J.C. \& Hillebrandt~W. 1995b, ApJ,  452, 779
 
\item Niemeyer~J.C. \& Woosley~S.E. 1997,  ApJ, 475, 740
 
\item Nomoto~K., Sugimoto~D., \& Neo~S. 1976, 
	Astrophys. Space Sci.,  39, L37
 
\item Nomoto~K., Thielemann~F.--K., \& Yokoi~K. 1984, 
	ApJ,  286, 644
 
\item N\o rgaard-Nielsen~H.U. et al. 1989, Nature,  339, 523
 
\item Perlmutter~S. et al. 1997, ApJ,  483, 565
 
\item Perlmutter~S. et al. 1999, ApJ,  517, 565

\item Phillips~M.M. et al. 1987, PASP, 99, 592
 
\item Phillips~M.M. 1993,  ApJ,  413, L105
 
\item Pskovskii~Yu.P. 1977, Sov.  Astronomy,  21, 675
        (translated from Russian: Astron. Zh. 1977,  54, 1188)
 
\item Pskovskii~Yu.P. 1984, Sov.  Astronomy,  28, 658
        (translated from Russian: Astron. Zh. 1984,  61, 1125)
 
\item Rees~M.J. 1998, in The Next Generation Space Telescope: Science
	Drivers and Technological Challenges, (astro-ph/9809029)
 
\item Reinecke~M., Hillebrandt~W., \& Niemeyer~J.C. 1999, 
        A\& A, in press
	(astro--ph/9812120)
 
\item Riess~A.G. et al. 1998, AJ, 116, 1009
\item Ruiz-Lapuente~P. 1997,
in proc. of Casablanca Winter School
{\it From Quantum Fluctuations to Cosmological Structures}, December 1996,
(ed. D.~Valls-Gabaud, M.~Hendry, P.~Molaro and K.~Chamcham),
ASP Conference Series, vol. 126, p. 207

\item  Ruiz-Lapuente~P., Blinnikov~S., Canal~R., Mendez~J., Sorokina~E.,
       Visco~A., \& Walton~N. 1999, in 
      {\it Cosmology with type Ia Supernovae}, proc. of the workshop held
at the University of Chicago, October 29 - 31, 1998 (ed. J.~Niemeyer \&
J.~Truran)

\item Sandage~G., Tammann~G.A. 1997, in {\it Critical Dialogues in
        Cosmology}, (ed.~N.~Turok), Singapore: World Scientific, p. 130
 
\item Schmidt~B.P. et al. 1998,	ApJ, 1998,  507, 46
 
\item  Sorokina E.I., Blinnikov, S.I.,
Ruiz-Lapuente P.,  \& Nomoto, K. 1999, in preparation

\item Suntzeff~N.B. et al. 1999, AJ, 117, 1175

\item Swartz~D.A., Sutherland~P.G., \& Harkness~R.P. 1995, 
        ApJ,  446, 766
 
\item Timmes~F.X. \& Woosley~S.E. 1992,  ApJ,
	 396, 649
 
\item   Umeda~H., Nomoto~K., Kobayashi~C., Hachisu~I., Kato~M. 1999,
   ApJ Letters, accepted (astro-ph/9906192)

\item Woosley~S.E. 1990a, in Supernovae (ed. Petschek~A.G.),
	Astron. Astrophys. Library, p.182
 
\item Woosley~S.E. 1990b, in Gamma-ray Line Astrophysics
        (ed. Durouchoux~P., Prantzos~N.), New York: Am. Inst. of Phys.,
	p.270
 
\item Woosley~S.E. 1997, in Thermonuclear Supernovae
        (ed. Ruis-Lapuente~P. et al.), Dordrecht: Kluwer Academic Publishers,
	p.313
 
\item Woosley~S.E. \& Weaver~T.A. 1986, 
	Ann. Rev.  Astron. Astrophys.,  24, 205
 
\item Woosley~S.E. \& Weaver~T.A. 1994, in Supernovae
        (ed. Audouze~J. et al.), Amsterdam: Elsevier Science Publishers,
	p.63
 
\item Zeldovich~Ya.B. \& Frank-Kamenetsky~D.A. 1938,  Acta 
	Physicochim. USSR,  9, 341
 
\end{description}
 
\newpage
 
\begin{table}
\caption{Parameters of the models}
\label{modpar}
\begin{center}
\begin{tabular}{llcccl}
\hline
\hline
model & burning mode & $M_{\rm WD}$(\Msun) & $E_{51}$
                           & $M_{{}^{56}{\rm Ni}}$(\Msun) & opacity \\
\hline
W7    & deflagration & 1.3775 & 1.20 & 0.60  & scattering \\
      &              &        &      &       & $\quad$ + absorption \\
W7a   & deflagration & 1.3775 & 1.20 & 0.60  & pure absorption \\
DD4   & delayed      & 1.3861 & 1.23 & 0.626 & scattering \\
      & detonation   &        &      &       & $\quad$ + absorption \\
\hline
\end{tabular}
\end{center}
\end{table}
 
\newpage
 
\begin{table}
\caption{Chemical composition of the models 
	(all masses are in units of \Msun)}
\label{modchem}
\begin{center}
\begin{tabular}{rcc}
\hline
\hline
            & W7 (W7a)            & DD4                 \\
\hline
\nifsx\hspace{3mm}      & $5.99\cdot10^{-1}$ & $6.26\cdot10^{-1}$ \\
He\hspace{3mm}          & $2.88\cdot10^{-3}$ &  0                  \\
C\hspace{3mm}           & $4.65\cdot10^{-2}$ & $3.28\cdot10^{-3}$ \\
O\hspace{3mm}           & $1.41\cdot10^{-1}$ & $1.01\cdot10^{-1}$ \\
Ne\hspace{3mm}          & $4.42\cdot10^{-3}$ & $1.20\cdot10^{-3}$ \\
Mg\hspace{3mm}          & $8.64\cdot10^{-3}$ & $8.97\cdot10^{-3}$ \\
Si\hspace{3mm}          & $1.52\cdot10^{-1}$ & $2.58\cdot10^{-1}$ \\
S\hspace{3mm}           & $8.62\cdot10^{-2}$ & $1.56\cdot10^{-1}$ \\
Ar\hspace{3mm}          & $1.60\cdot10^{-2}$ & $3.47\cdot10^{-2}$ \\
Ca\hspace{3mm}          & $1.24\cdot10^{-2}$ & $3.75\cdot10^{-2}$ \\
Fe\hspace{3mm}          & $1.45\cdot10^{-1}$ & $1.59\cdot10^{-1}$ \\
${}^{58}$Ni\hspace{3mm} & $1.63\cdot10^{-1}$ &  0                  \\
\hline
\end{tabular}
\end{center}
\end{table}
 
\begin{table}
\caption{Rise time to maximum light in different colors}
\label{rise}
\begin{center}
\begin{tabular}{lccccccc}
\hline
\hline
model & $t_{\rm rise}^B$ & $M_{\rm max}^B$          & $t_{\rm rise}^V$ 
      & $M_{\rm max}^V$  & $t_{\rm rise}^{\rm bol}$ & $M_{\rm max}^{\rm bol}$
      & $\log L_{\rm max}^{\rm bol}$ \\
\hline
DD4   & 17.1 & --18.91 & 22.1 & --19.03 & 9.4 & --19.36 & 43.23 \\
W7    & 16.0 & --18.71 & 20.5 & --18.82 & 8.6 & --19.38 & 43.24 \\
W7a   & 11.9 & --18.97 & 13.9 & --19.18 & 8.9 & --19.42 & 43.25 \\
\hline
\end{tabular}
\end{center}
\end{table}

\newpage
 
 
\newpage
 
\newpage
 
\newpage
 
\begin{figure}
\plotone{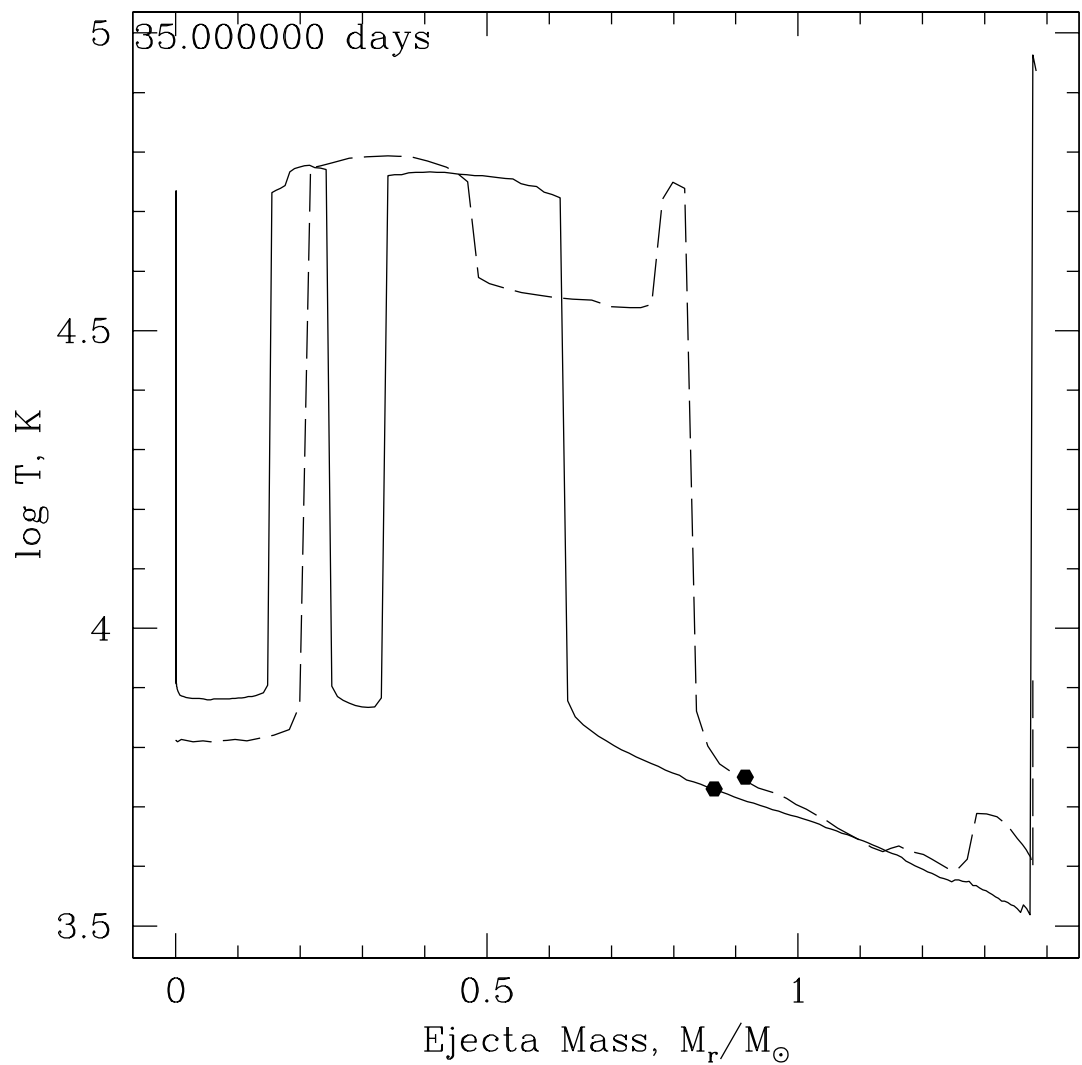}
\caption{The temperature distribution  for
DD4 (solid) and W7 (dashed) on 35th day after the explosion. 
The solid dots show the locations where the
optical depth measured from the surface reaches the value of 2/3.}
\label{tmpr}
\end{figure}
 
\newpage
 
\end{document}